\documentclass{article}
\usepackage[american]{babel}
\usepackage[latin1]{inputenc}
\usepackage{graphicx}

\begin{document}

\title{Bayesian Updating Rules in Continuous Opinion Dynamics Models}


\author{Andr\'e C. R. Martins\\
GRIFE -EACH\\  
Universidade de S\~ao Paulo}

\maketitle


\begin{abstract}

 In this article, I investigate the use of Bayesian updating rules applied to modeling social agents in the case of continuos opinions models. Given another agent statement about the continuous value of a variable $x$, we will see that interesting dynamics emerge when an agent assigns a likelihood to that value that is a mixture of a Gaussian and a Uniform distribution. This represents the idea the other agent might have no idea about what he is talking about. The effect of updating only the first moments of the distribution will be studied. and we will see that this generates results similar to those of the Bounded Confidence models. By also updating the second moment, several different opinions always survive in the long run. However, depending on the probability of error and initial uncertainty, those opinions might be clustered around a central value.

\vspace{0.2cm}

\end{abstract}

\section{Introduction}

Opinion Dynamics~\cite{bordognaalbano07,castellanoetal07,galametal82,galammoscovici91,sznajd00,stauffer03a,deffuantetal00,hegselmannkrause02} aims to understand how simple interactions between artificial agents can describe the aspects we observe on the opinions of social groups. Typical models describe opinions as either discrete~\cite{galametal82,galammoscovici91,sznajd00,stauffer03a} or continuous~\cite{deffuantetal00,hegselmannkrause02,bennaimetal03,weisbuchetal05,deffuant06} variables. Under those models, the interacting agents observe the complete opinion of each of the other agents they interact with. However, this is not necessarily the case and agents might not express their opinions fully, due to verbalization problems\cite{urbig03}.

In an earlier work, I have explore the use of Bayesian updating rules in the context of binary expression of choices, with a continuous underlying probability associated to each choice, the Continuous Opinions and Discrete Actions (CODA) model~\cite{martins08a,martins08b}. That is, the verbalization in CODA is limited to two choices. Each agent had a continuous opinion, a probability that one of two choices was the best one, but only observed the discrete choices of its neighbors. That allowed the modelling of the emergence of extremism, even when no extremist agents were observed initially. 

In this paper, the application of Bayesian rules to a purely continuous problem, with no verbalization problems, will be studies. The objective is to compare the results we obtain when using Bayesian rules, as in the CODA model, with the results of traditional continuous opinion models. Here, agents tell their continous estimate about the value of a variable $x$ to each other and change their minds according to update rules obtained as approximations to the Bayesian inference problem. By comparing the results of the Bayesian update with traditional continuous models, we will see that bounded confidence models correspond to the case where only the first moment of the opinion distribution is updated. In that case, we will see that Bayesian models provide basically the same qualitative features of the bounded confidence models and that one can recover the bounded confidence update rule as an approximation. By also updating the second moment, new features of opinion evoltuion will be observed and we will see that, in the long run, the agents tend to become very certain about their own choices.

\section{Bayesian Update for Continuous Variables}\label{sec:contionuousupdate}

Draw two agents at random and let them exchange their views on the value of a continuous variable $x$, where $0\leq x\leq 1$. This might represent the case where agents speak about their inner probabilities or when they try to reach a consensus about the value of some continuous parameter, rescaled to that interval. In bounded confidence models~\cite{deffuantetal00,hegselmannkrause02}, this problem is represented by each agent $i$ having a continuous opinion $x_i$ about the value of a parameter $\theta$. As they interact, they change their opinions towards that of the other agent (or agents), as long as the difference between their opinions, $|x_i-x_j|$ is not above a certain threshold $\epsilon$.

To obtain an approximation to a Bayesian inference of this problem, we need to express the initial opinion of each agent as a prior probability distribution $f_i(\theta)$, such that $E_i[ \theta]=x_i$, where $E_i$ means the expected value agent $i$ associates with $\theta$, given its probabilistic opinion $f_i$. The function $f_i(\theta)$ will, of course, be altered as the agent observes the average estimates $x_j$ of other agents, leading to a new, posterior distribution, $f(\theta|x_j)$. One of the simplest possible choices is to model the initial prior opinion as a Normal distribution. That means that the opinion $f_i(\theta)$ of agent $i$ should include not only an average $x_i$, but also an uncertainty $\sigma_i$, to be used as the standard deviation of the Normal opinion. As an agent observes that his neighbor average estimate is $x_j$, it will need a likelihood funtion that models how likely it is for the neighbor to have that average estimate as a function of the true value of the parameter $\theta$, that is, an estimate for $f(x_j|\theta)$. 

A true likelihood would have to model how the value of $\theta$ influences $x_j$. Since $x_j$ is also influenced by $j$ interactions with other agents, that fact, in a complete model, would also have to be included. And that means agent $i$ would have to model how agent $j$ models every other agent and how many interactions $j$ has had so far, including his model of how $i$ reasons. However, although correct, from a Bayesian point of view, this regress might not be a reasonable description of real agents. And one should not forget the goal of looking for simple models. Therefore, simpler likelihoods are needed. 

The first simple idea is, of course, to use a Normal distribution for the likelihood, with an average at $x_j$. One extra level of detailing is still needed to define the likelihood and that is how close $x_j$ is likely to be to the true value of $\theta$. A natural candidate for that role is the uncertainty $\sigma_j$ of neighbor $j$. Assuming $i$ knows the value of $\sigma_j$, this likelihood $N(x_j,\sigma_{j}^{2})$ will change $f_i(\theta)$ to $f(\theta|x_j,\sigma_{j})$ that is also a Normal distribution. The average opinion of agent $i$ becomes $x'_i$, given by the weighted average of $x_i$ and $x_j$

\begin{equation}\label{eq:noerror}
x'_i=\frac{\frac{x_i}{\sigma_{i}^{2}}+\frac{x_j}{\sigma_{j}^{2}}}{\frac{1}{\sigma_{i}^{2}}+\frac{1}{\sigma_{j}^{2}}}.
\end{equation}
If one assumes that agents do not share information about their uncertainty, as we will, from this point on, a reasonable assumption for agent $i$ is that $\sigma_i=\sigma_j$. That is, agent $i$ assumes that the neighbor knowledge is as good as its own. In this case, Equation~\ref{eq:noerror} becomes a simple average between $x_i$ and $x_j$.

Unfortunately, the model that we get from this is trivial, even when  $\sigma_i\neq\sigma_j$. If one assumes no social structure and that both interacting agents update their opinions, as in the bounded confidence models, it is easy to see that this description will converge to a single value in the long run. The introduction of a social structure will not solve the problem, since no true boundaries will appear. In the CODA model, the dissention and the opposing extremist opinions appear because when an agent observes a neighbor that disagrees, that agent will move only a fixed step towards the opposite direction. For Equation~\ref{eq:noerror}, the further the opinions are, the larger the movement each agent makes towards each other and, therefore, there is no strenghtening of the domains. Two agents who have far opinions have a stronger influence on each other than agents with similar opinions. This needs to be corrected, if we mean to model the way real people think.

\section{Fixed Uncertainty}

In order to get a more interesting and realistic model, a mechanism similar to those of the bounded confidence models is required. In those models, agents $i$ and $j$ would only change their opinions when their opinions were not too distant, that is, when $|x_i-x_j|$ is smaller than a threshold $\epsilon$. It is this lack of influence of too distant opinions that allows different points of view to survive in the long run. Therefore, it would be interesting to develop a similar mechanism.

\begin{figure}[bbbh]
 \includegraphics[width=0.8\textwidth]{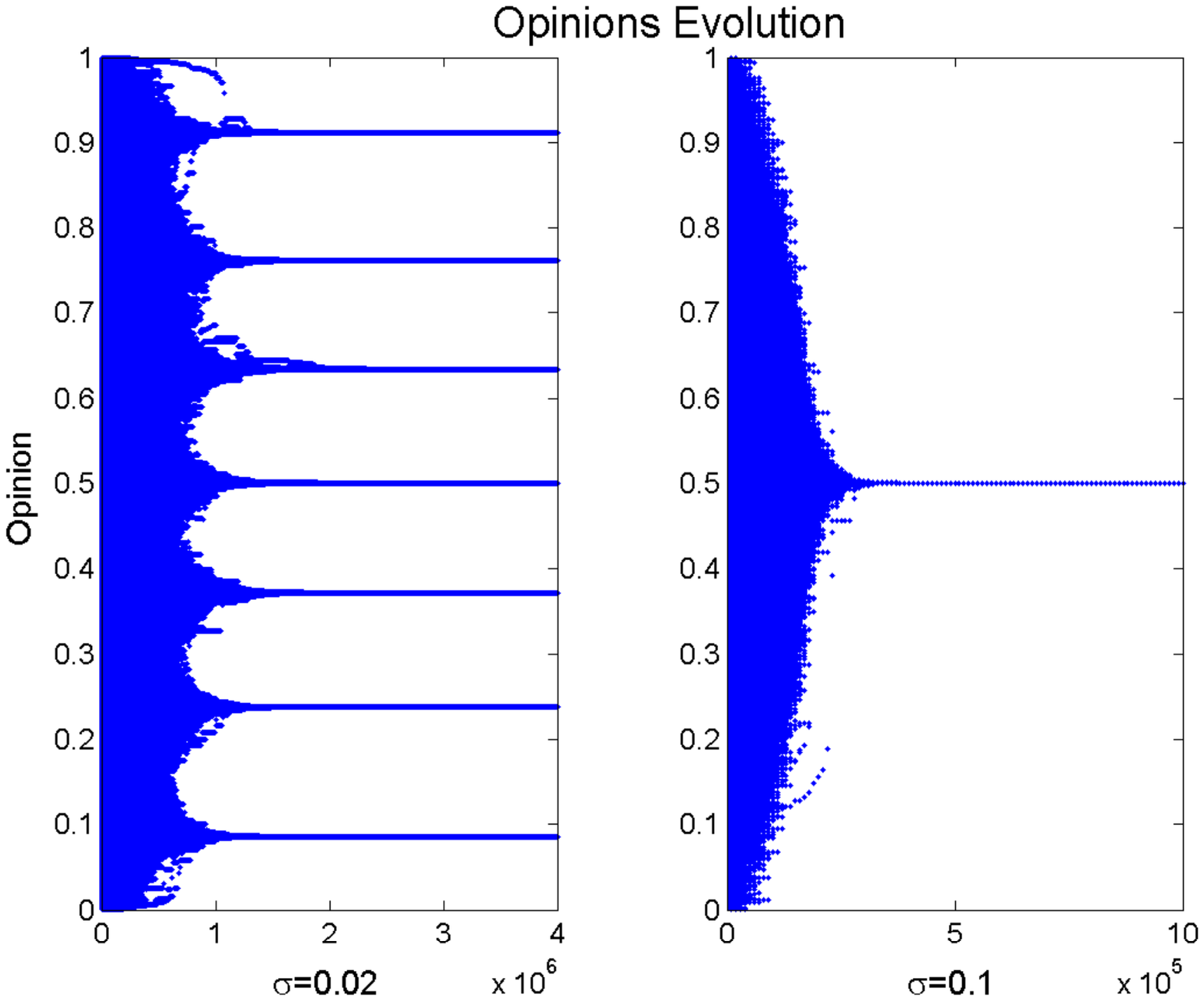}
 \caption{Time evolution of opinions for the fixed variance case, for 10,000 agents.}
 \label{fig:timeevolcv}
 \end{figure} 
 
We can understand that the reason why agents with very different opinions fail to interact is related to the fact that agents might not trust someone whose opinion is too different of their own. For a Bayesian model, there is a natural way to express the same concern. This can be achieved by introducing a probability $p$ that the other agent actually knows something about $\theta$. When that happens, the likelihood can be modelled as the Normal distribution of the the non-error case. But there will be a $1-p$ chance that the other agent has no real information about $\theta$. In this case, instead of a Normal likelihood, peaked around the opinion $x_j$ of the other agent, it would make more sense to choose a non-informative likelihood. Therefore, we can use the likelihood
\begin{equation}\label{eq:likelihooddecept}
f(x_j|\theta) = p N(\theta,\sigma_{j}^{2}) + (1-p) U(0,1),
\end{equation}
where $N(\theta,\sigma_{j}^{2})$ represents the Normal distribution centered around $\theta$ and $U(0,1)$ represents an uniform distribution between 0 and 1.

We are, as stated before, assuming that only information about $x_i$ is exchanged. Therefore, the values of $\sigma_i$ need to be guessed. Again, we can make the reasonable assumption that all agents have similar uncertainty. From now on, whenever an agent interacts with another, it will assume that the uncertainty associated with the Normal part of the likelihood of the other agent is equal to its own uncertainty. With this, we can multiply the likelihood in Equation~\ref{eq:likelihooddecept} by the Normal prior discussed before, in order to obtain a posterior distribution. Each term in Equation~\ref{eq:likelihooddecept} contributes as one Normal term in the posterior, that is a mixture of two Normals. That is,
\begin{equation}\label{eq:posterior}
f(\theta|x_j)\propto p e^{-\frac{1}{2\sigma_{i}^{2}}[(\theta-x_i)^2+(x_i-\theta)^2]}+(1-p) e^{-\frac{(x_i-x_j)^2}{2\sigma_{i}^{2}}}
\end{equation}
It is important to notice that, since we are actually using hierarchical Bayesian models, the normalization constants have to be used and can no longer be ignored in teh calculations.
The second term is a Normal distribution with an average equal to the prior average $x_i$; it is equivalent to the result where no change happens, because the agent believes the other agent might knows nothing. The other Normal corresponds to the interaction case, where the agents opinions tend to each other. After rearranging the terms in the exponentials, we can calculate the expected value of $ \theta$ in Equation~\ref{eq:posterior}, $E_i[ \theta]$, and we see that $x_i(t)$ is transformed to
\begin{equation}\label{eq:averagedecept}
x_i(t+1)=p^*\frac{x_i(t)+x_j(t)}{2}+(1-p^*)x_i(t)
\end{equation}
where
\begin{equation}\label{eq:posteriorp}
p^* = \frac{p\frac{1}{\sqrt{2\pi}\sigma_i} e^{-\frac{(x_i(t)-x_j(t))^2}{2\sigma_{i}^{2}}} }{p\frac{1}{\sqrt{2\pi}\sigma_i} e^{-\frac{(x_i(t)-x_j(t))^2}{2\sigma_{i}^{2}}} +(1-p)}.
\end{equation}

 Notice that the posterior average is an average between $\frac{x_i(t)+x_j(t)}{2}$, the result one would expect if there were no agents who know nothing about $\theta$, and just $x_i(t)$, that corresponds to the possibility that the other agent knows nothing. However, the weight to each possibility is not given by $p$, but by an altered value $p^*$. If $x_i(t)-x_j(t)$ is small when compared to $\sqrt{2}\sigma_{i}$, the first term in the denominator of Equation~\ref{eq:posteriorp} will be larger than the second term and $p^*$ will become closer to 1 than $p$, meaning that agents with similar minds will tend to the average between their opinions. On the other hand, as  $x_i(t)-x_j(t)$ becomes larger, the numerator becomes smaller and, for $x_i(t)-x_j(t)$ large enough, it will tend to zero, while the denomiator tends to a constant value ($1-p$). Under these circumstances, the agents $i$ and $j$ will influence each other very weakly. Some influence will always exist, unlike the bounded confidence models, but it can be negligible.
 
\begin{figure}[bbbh]
 \includegraphics[width=0.8\textwidth]{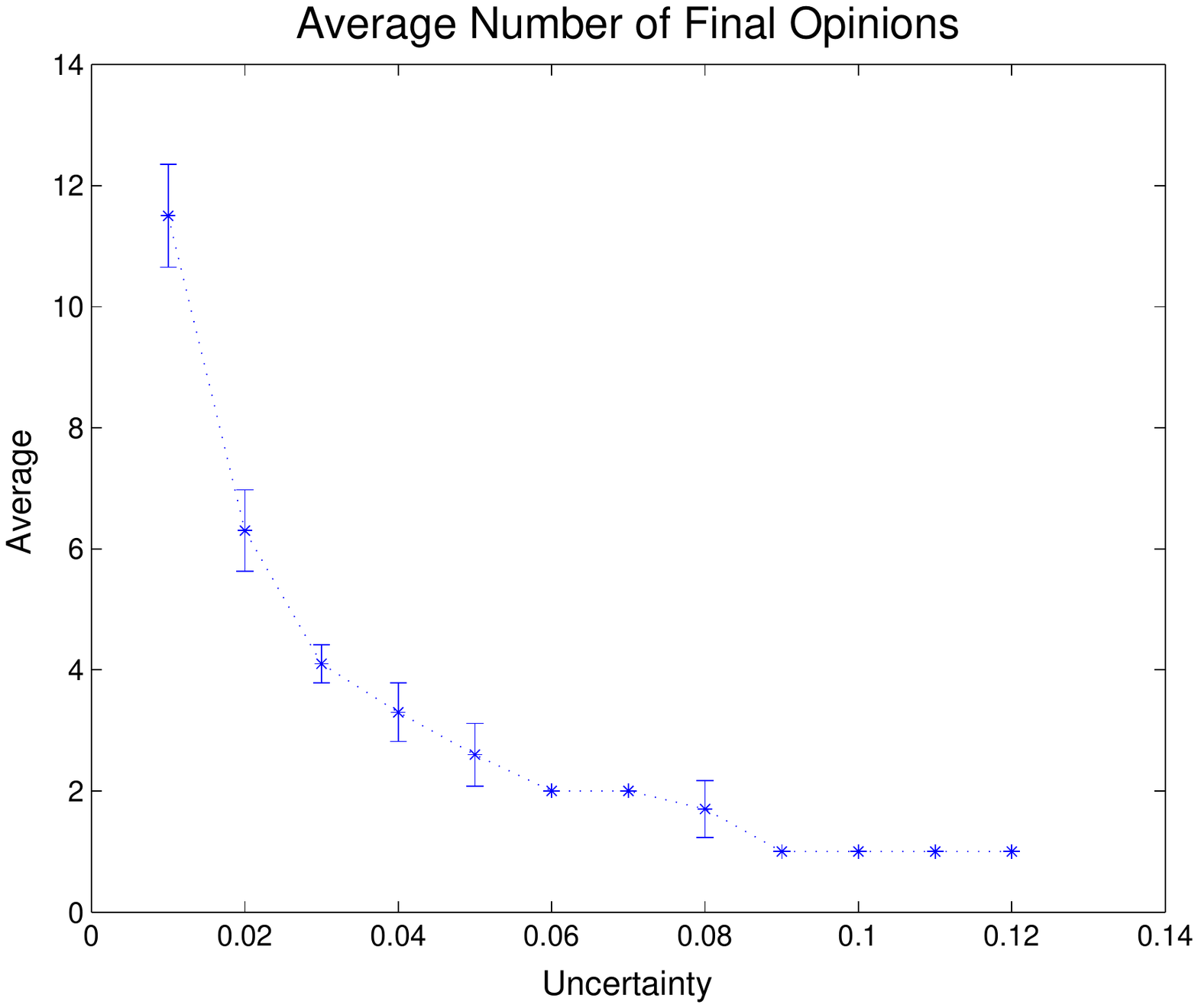}
 \caption{Number of final opinions as a function of  $\sigma$. Each point corresponds to the average  number of opinions for 10 realizations of the problem and the error bars correspond to the standard deviation of that number.}
 \label{fig:finalop}
 \end{figure} 

 A number of simulations were performed where each agent only updates its average opinion about $\theta$, that is, its value $x_i$ and nothing else. Simulations were run for $p=0.7$ as well as $p=0.99$, and only minor quantitative differences were observed. This seems to suggest that the exact value of $p$ is not so important. Apparently, $p$ has an impact in the velocity agents change their minds. Figure~\ref{fig:timeevolcv} shows the evolution of the opinions of 10,000 agents when all of them keep their uncertainty constant, for $p=0.7$. It shows two cases, for $\sigma=0.02$, corresponding to a population that is reasonably certain about their opinions, and $\sigma=0.1$, that represents a larger uncertainty. We can see that, in the first case, the final state converges to seven different opinions and that this final state is stable. For a larger uncertainty, on the other hand, the agents are able to reach consensus. Those results are qualitatively the same as that previously observed in the bounded confidence models~\cite{deffuantetal00,hegselmannkrause02}.  
 
In the bounded confidence case, the threshold $\epsilon$ was the parameter that controlled how many opinions survived in the end. Here, the initial uncertainty $\sigma$ plays the role of the threshold. The use of Normal distribution as basis for the interaction was studied by Deffuant~\cite{deffuant06}, in the Gaussian Bounded Confidence model, but the actual details of the update rule in that study are different from Equation \label{eq:averagedecept}. For small values of the uncertainty, $p^*$ will tend to zero even for small distances between the opinions, while a large $\sigma$ means that distant opinions can still influence each other. Figure~\ref{fig:finalop} shows the average number of final observed opinions as a function of the uncertainty $\sigma$. Notice that for $\sigma\geq 0.1$, only one final opinion was observed in all cases; as $\sigma\rightarrow 0$, the surviving number of opinions increases fast.

\section{Evolving Uncertainty}\label{sec:uncertainchange}
 
It is interesting to notice that one simple next step towards a more complex agent is to update one more variable as the agents interact. After updating the first momentum of the distribution for $\theta$, it is natural to ask what happens what happens when the second momentum is also updated, that is, when the uncertainty associated with that value also changes. It is easy to see that Equation~\ref{eq:posterior} is still valid, since it is the correct inference about the first momentum. The only difference now is that the agents will also update their uncertainty to match the variance of the posterior and this will influence the future evolution of the first momentum. A similar change in the threshold $\epsilon$ was implemented by Weisbuch et al~\cite{weisbuchetal05} in the context of bounded confidence models.

\begin{figure}[bbbh]
 \includegraphics[width=0.8\textwidth]{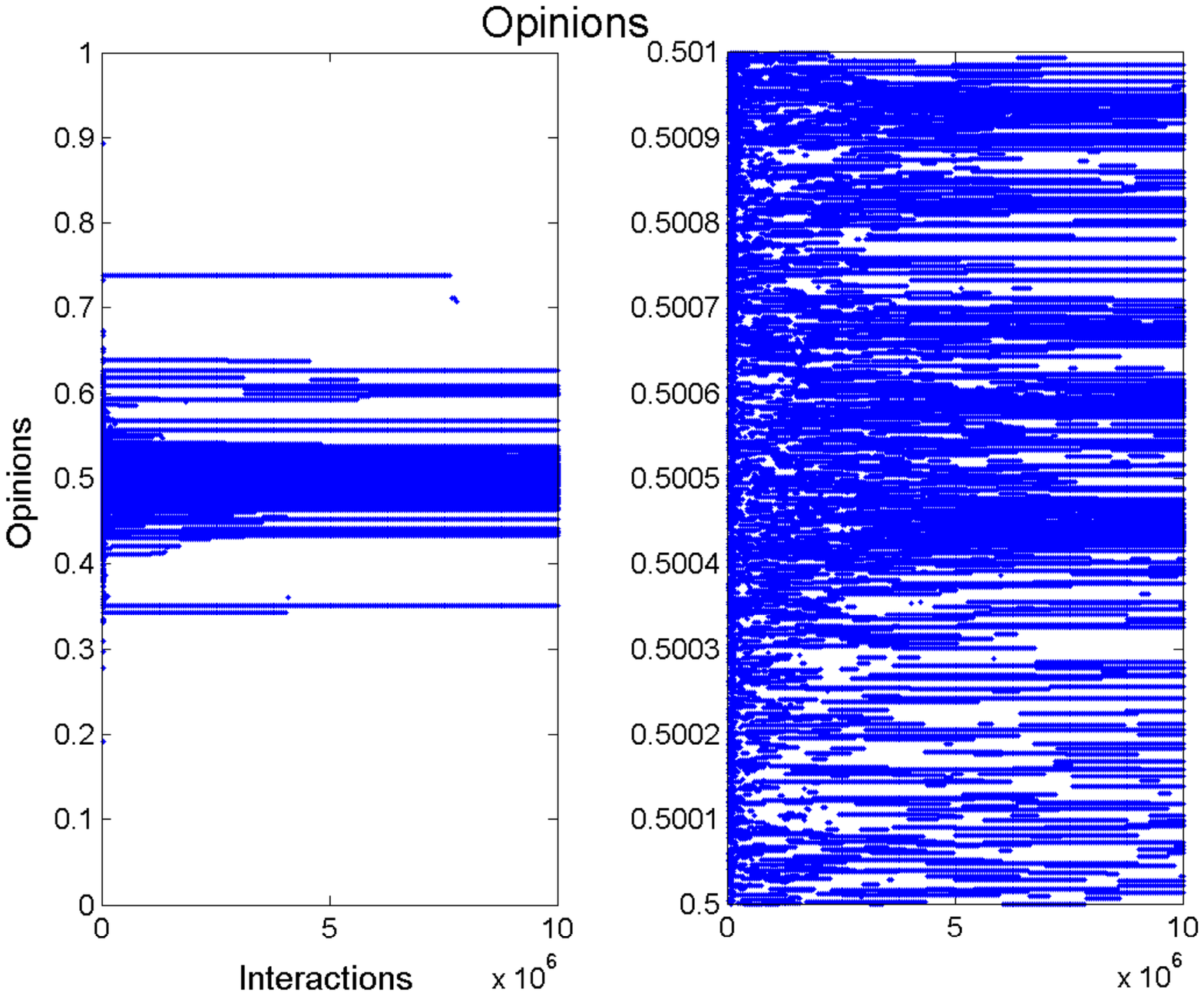}
 \caption{Time evolution of the opinions in the changing variance case. The case shown corresponds to 10,000 agents, after a total of 1,000,000 interactions, with initial uncertainty of 0.5.}
 \label{fig:timeevol05}
 \end{figure} 
 
By estimating $\sigma_{i}^{2}=E[ \theta^2]-E[ \theta]^2$ from the posterior in Equation~\ref{eq:posterior}, the new uncertainty $\sigma_i(t+1)$ becomes, after some straightforward calculations,
\begin{equation}\label{eq:varupdate}
\sigma_{i}^{2}(t+1)=\sigma_{i}^{2}(t)\left( 1-\frac{p^*}{2} \right) + p^*(1-p^*) \left(\frac{x_i(t)-x_j(t)}{2}\right)^2.
\end{equation}

Figure~\ref{fig:timeevol05} shows the time evolution of the opinion of the agents when they also update their uncertainty. Here we observe a different dynamics than those we have observed for constant uncertainty. The agents tend to the central opinion. However, while most of the agents end near 0.5 (in that specific run, the final average of the opinions of the agents was 0.4977 and the standard deviation, 0.0072), we still observe a few agents who final opinion is far from the average value. As a matter of fact, the smallest observed value of $p$ in that run was $p=0.2627$ and the largest, $p=0.5994$. That is, in the end, even one opinion more than 32 standard deviations from the average one survived.

\begin{figure}[bbbh]
 \includegraphics[width=0.8\textwidth]{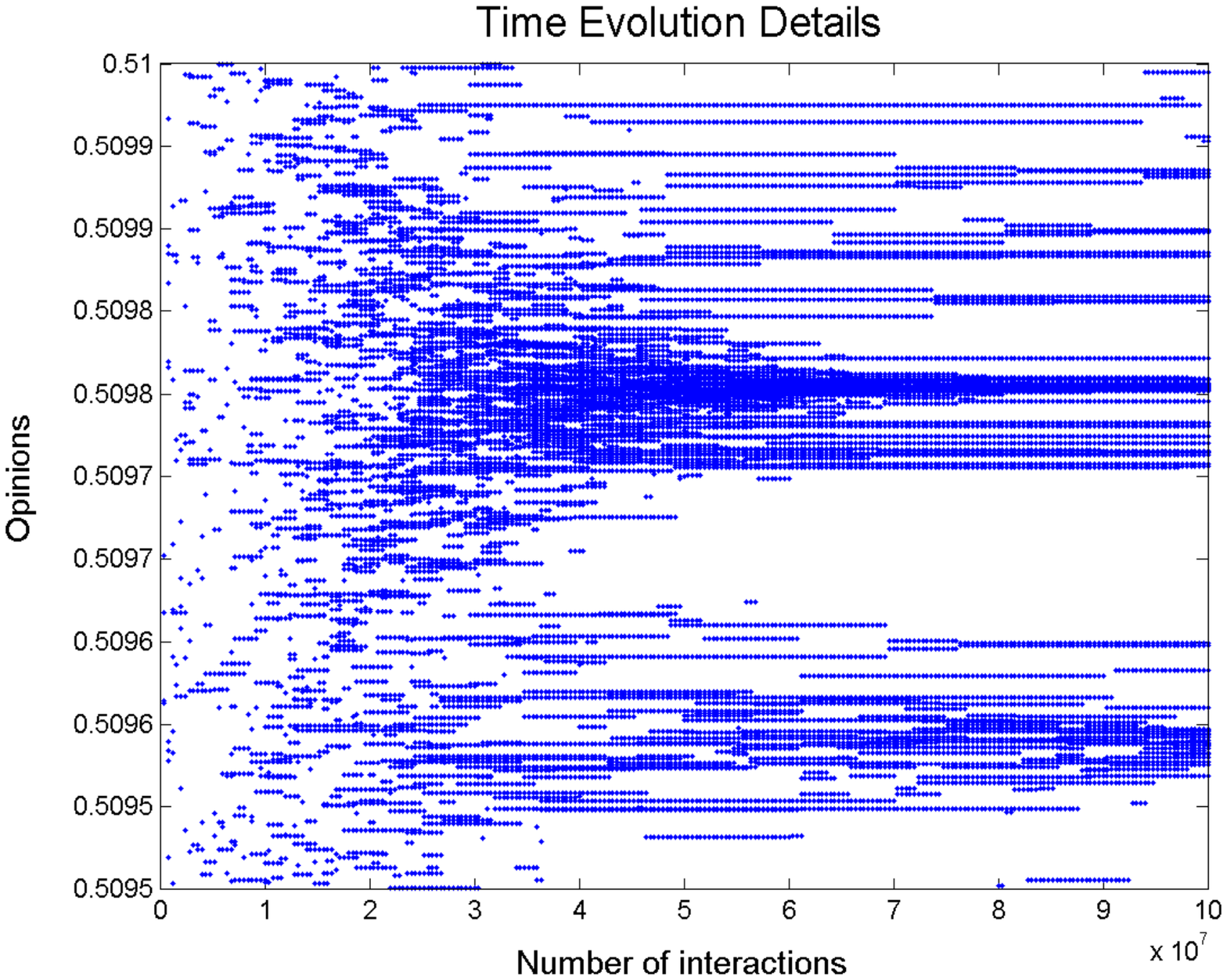}
 \caption{Fine details of the time evolution of the opinions in the changing variance case. The case shown corresponds to 20,000 agents, after a total of 10,000,000 interactions, with initial variance of 0.01.}
 \label{fig:timeevol}
 \end{figure} 
 
Another interesting characteristic of this model can be observed in the second panel of Figure~\ref{fig:timeevol05}, where a zoom to the region of opinions between 0.5 and 0.501 is shown. At a very small scale, the final opinions do not actually converge, but they separate into a myriad of different, although very close, values. This effect can be better observed in Figure~\ref{fig:timeevol}, where 20,000 agents were left to interact longer (10,000,000 total interactions), so that this effect could be better visualized. Only the region betweem 0.5095 and 0.51 is shown, in order to see the fine details. While some agents still change their opinions at this level of detail, the changes become rarer and smaller with time. During the beginning of the simulation, the opinions changed wildly and no straight lines were observed. Around 3 to  million interactions after the start, each opinion become more stable and this process goes on.

What is really happening is that Equation~\ref{eq:varupdate} leads to a decrease in the uncertainty with time. This effect was partially described in Weisbuch model~\cite{weisbuchetal05} by the fact that agents with smaller thresholds would change their threshold less than agents with larger thresholds. However, in that model, the threshold would never end smaller than the smallest of the agents thresholds. Here, the interaction will cause initially identical uncertainties to become different, due to the dynamics of the model, and they will also become much smaller than any of the uncertainties introduced at $t=0$.

This dimishing tendency is easy to see by analysing the two terms of Equation~\ref{eq:varupdate}. Notice that the first term causes the uncertainty to either remain the same (when $p^*=0$) or to decrease. The minimum value for one interaction is half the previous one, when $p^*\approx 1$. The second term could, in principle, make the uncertainty becomes larger, since it is always a positive contribution. However, when $p^*$ is close to 1 or 0, the second term becomes close to zero. For intermediary values of $p^*$, it actually slows down the diminishing of $\sigma_i$, but its effect is not strong enough and, in the long run, $\sigma_i\rightarrow 0$. In the simulation run shown in Figure~\ref{fig:timeevol}, the initial uncertainty was 0.1 and, at the end of the simulation, the agents had an average uncertainty of $\bar{\sigma_i}=1.59\cdot 10^{-5}$. This means that only very close opinions will be able to influence each other and, as they do, $\sigma_i$ will become even smaller. In the $n\rightarrow\infty$ limit, of infinite agents, there should always be a close enough agent so that some change might still happen, but, those changes will be smaller and smaller.

\begin{figure}[bbbh]
 \includegraphics[width=0.8\textwidth]{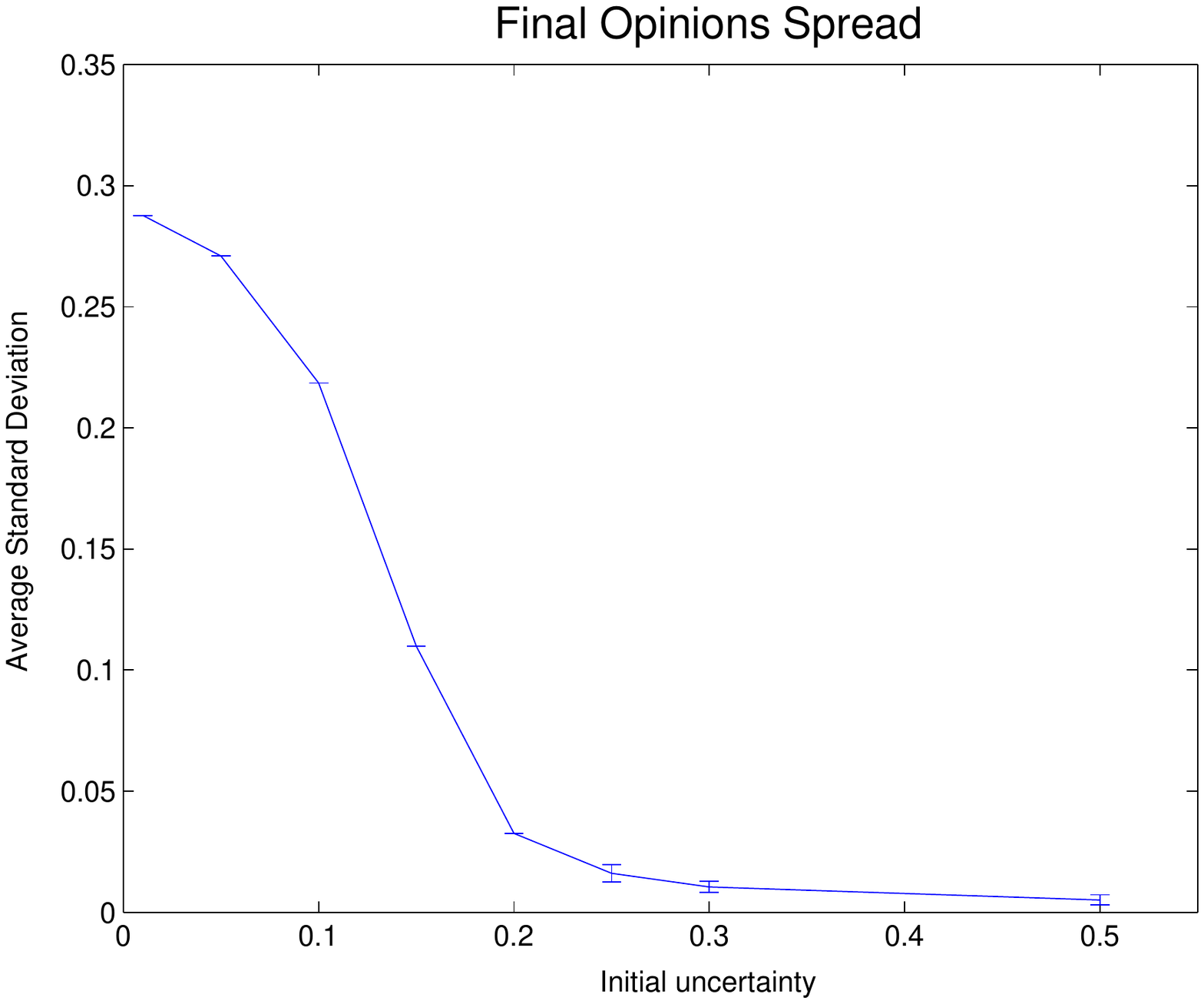}
 \caption{Standard deviation of the final opinions as a function of initial uncertainty. For an initial uncertainty of zero, the distribution is uniform and, therefore, the standard deviation is $1/\sqrt{12}\approx 0.289$}
 \label{fig:varupdatespread}
 \end{figure} 
 
While the final opinions become rigid, it is interesting to measure the spread of the final state. Figure~\ref{fig:varupdatespread} shows the standard deviation of the final opinions for different values of the initial uncertainty. We can see that, as the initial uncertainty becomes smaller and closer to zero, the final opinions tend not to change and the initial uniform standard deviation ($1/\sqrt{12}\approx 0.289$) is preserved. For larger amounts of initial uncertainty, however, a tendency to more central opinions is observed, indicated by the fast decrease of the standard deviation of the final opinions. However, the final standard deviation seems to only tend to zero when the initial uncertainty tends to infinity, and that was confirmed by other simulations with a larger initial value for $\sigma_i$. The problem is related with the fact that the uncertainty of the agents decrease fast when they actually change their opinions and, therefore, a state with small uncertainty is soon achieved. That state prevents the final standard deviation of decreasing further and it seems it never becomes really zero. It can still be very small, of course, indicating a final state where the population have close, but not identical, opinions.

Finally, it is interesting to notice that the behavior of the standard deviation shown in Figure~\ref{fig:varupdatespread} do not depend or depend only very weakly on the number of agents. A simulation with as little as $n=200$ agents obtained not only the same general shape of the curve, but, basically, the same final values with only very small differences, whithin the error bars.

\section{Conclusions}

We have seen that Bayesian rules can also provide a way to model the change in the opinion of agents when the information that is exchanged between the agents is continuous. If we approximate Equation~\ref{eq:posteriorp} by a step function, with values of 0 or 1, depending on how close $x_i$ and $x_i$ are, we recover the results of bounded confidence models. That is, the bounded confidence models can be seen an approximation to a fixed uncertainty model described here. We have seen that the same qualitative description is obtained and that the initial uncertainty $\sigma$ is the parameter that is equivalent to the threshold $\epsilon$ of the bounded confidence models. We have also shown that, by updating the second momentum, we get a model that is analogous to the update of the threshold in Weisbuch et al~\cite{weisbuchetal05}. This makes it possible to clearly identify the threshold as an approximation to the second momentum of the probability distribution an agent might have about the variable that is discussed. With the current exercise, one of the good qualities of using Bayesian rules became evident. By understanding what approximations we have to do in order to obtain the traditional results of bounded confidence models, we gain a better understanding of those models as well as a natural way to propose more complex models. This can be achieved by avoiding one or more approximations.

 We should notice that, even when updating the uncertainty, we are still far from a correct inference, since the true posterior distribution is a mixture distribution but the agents only record the two first moments. By forgetting everything but the two momenta from one interaction to the next one, a Normal distribution is actually the choice that maximizes entropy. Therefore, it makes sense to use it to approximate the inference proccess, but it is important to notice that the prior used in the next interaction is not the posterior of the previous one. Still, this new, a little more sophisticated, model showed interesting properties. In particular, we have seen that the agents tend to become more certain about their opinions with time and more interactions. After a while, even the opinions of other agents with close opinions are disregarded as wrong, since each one is very certain about their own ideas. Therefore, it is possible to have some convergence of the population to a common value, but the agents become stubborn and less capable of learning.

Two different kinds of extremism can be discussed this way. One, as observed in the CODA model, correspond to a society where the agents have wildly different opinions. This case is similar to what we have observed here when the initial uncertainty is very small, since very different opinions can survive at the end. But we have also observed a second, less dangerous kind, when most agents have similar opinions, but are no longer capable of influencing each other after a while. Since the social effect is basically agreement, it seems this is better described as stuborness rather than true extremism and true extremism should really be identified as extreme values of $x_i$. In the evolving uncertainty model, what we have observed is that, with time, agents may converge to close opinions, but they become more and more stubborn about their own positions.

It is important to stress once more that these agents here are not Bayesian agents and don't have any higher cognitive abilities. The Bayesian rules were actually used only in order to find reasonable rules of interactions. Once those are found, the agents follow them in a dumb way. Modelling rational agents can be an interesting project, but that was not the approach of this paper. We have seen that there is another less complex option and simple rules can be a good choice when we are interested only in the description of the population and not the details of specific agents.

\section{Acknowledgement}
The author would like to thank Funda\c{c}\~ao de Ampara \`a Pesquisa do Estado de S\~ao Paulo (FAPESP), for the support to this work, under grant 2008/00383-9.

\bibliographystyle{plain}
\bibliography{extremism}

\end{document}